\documentclass[aip,jcp,reprint]{revtex4-1}

\usepackage{dcolumn}
\usepackage{bm}
\usepackage{graphicx}
\usepackage{amsmath, amsthm, amssymb, amsbsy}
\usepackage{color}
\usepackage{ulem}
\usepackage{hyperref}

\begin{document}

\title{Molecular Dynamics Simulation Study of Nonconcatenated Ring Polymers in a Melt: I. Statics}

\author{Jonathan D. Halverson}
\affiliation{Max Planck Institute for Polymer Research, Ackermannweg 10, 55128 Mainz, Germany}

\author{Won Bo Lee}
\affiliation{Max Planck Institute for Polymer Research, Ackermannweg 10, 55128 Mainz, Germany}
\affiliation{Department of Chemical and Biomolecular Engineering, Sogang University, Shinsu-dong 1, Mapo-gu, Seoul, Korea}

\author{Gary S. Grest}
\affiliation{Sandia National Laboratories, Albuquerque, NM 87185, USA}

\author{Alexander Y. Grosberg}
\affiliation{Department of Physics, New York University, 4 Washington Place, New York, NY 10003, USA}

\author{Kurt Kremer}
\altaffiliation{The following article has been accepted by \textit{The Journal of Chemical Physics}. After it is published, it will be found at \url{http://jcp.aip.org}. Corresponding author e-mail: kremer@mpip-mainz.mpg.de}
\affiliation{Max Planck Institute for Polymer Research, Ackermannweg 10, 55128 Mainz, Germany}

\date{\today}

\begin{abstract}
Molecular dynamics simulations were conducted to investigate the
structural properties of melts of nonconcatenated ring polymers and compared to melts of linear polymers. The longest rings were composed of $N=1600$ monomers per chain which corresponds to roughly 57 entanglement lengths for comparable linear polymers. For the rings, the radius of gyration squared, $\langle R_g^2 \rangle$, was found to scale as $N^{4/5}$ for an intermediate regime and $N^{2/3}$ for the larger rings indicating an overall conformation of a crumpled globule. However, almost all beads of the rings are ``surface beads'' interacting with beads of other rings, a result also in agreement with a primitive path analysis performed in the next paper\cite{halverson_part2}. Details of the internal conformational properties of the ring and linear polymers as well as their packing are analyzed and compared to current theoretical models. [DOI: 10.1063/1.3587137]
\end{abstract}

\maketitle

\section{Introduction}
\label{sec:1}

Understanding the static and dynamic properties of ring polymer
melts is one of the remaining challenges in polymer science. Unlike linear polymers,
the topological constraints for ring or cyclic polymers are permanent and this affects both their
static and dynamic properties. For linear polymers the topological
constraints imposed by the non-crossability of the chains
(entanglements) force each chain to diffuse along its own
coarse-grained backbone, the so-called primitive path, and this is well
described by the reptation model of de Gennes and Edwards. For
branched systems strands have to fold back in order to find a new
conformation without crossing any other chain, resulting in an
exponential growth of the longest relaxation time due to the entropy
barrier of ${\cal O}(\mathrm{strand~length})$ between different
states\cite{deGennes71,deGennes79,Doi86}. A number of simulations and
experimental results confirm this
concept\cite{Tobolsky,Onogi70,Berry68,Casale71,Odani72,Kurt90,Binder95,McLeish02,Everaers04}.
For both linear and branched polymer melts, it is the free chain
ends that make the known relaxation
mechanisms possible. However, in the case of rings there are no
free ends. This raises a number of unanswered questions regarding
the motion and stress relaxation of ring polymer melts based on
their different conformational properties, which will be discussed
in the subsequent paper\cite{halverson_part2}.

Understanding a melt of nonconcatenated and unknotted rings is also
closely related to the problem of the compact globular state of an
unknotted loop. Grosberg et al. \cite{Grosberg_Nechaev_Shakhnovich_1988} were
the first to hypothesize that the equilibrium state of a
compact unknotted loop is the so-called crumpled
globule, in which each subchain is condensed in itself, and,
therefore, the polymer backbone is self-similar with a fractal dimension
of 3.  The idea behind this prediction is that two pieces 
of the chain, each of them crumpled, act pretty much like two nonconcatenated rings, and the
latter system obviously experiences at least some volume exclusion
for topological reasons (on this subject see recent work
\cite{Bohn_Heerman_(from_Heidelberg)_Topological_Loop_Repulsion} and
references therein). A direct test of this prediction was first made
by Lua et al. \cite{Lua_Borovinskiy_Grosberg_POLYMER_2004} by modeling the
closed Hamiltonian loops on the compact domains of a
cubic lattice. Some
evidence of segregation between globules on the chain was observed,
but overall the results were inconclusive because the simulated
chains were too short or the statistics were too poor.

Squeezing topologically constrained rings against each other in a
melt, or squeezing one unknotted ring against itself in a restricted
volume, is also a problem of great potential significance in
application to the DNA organization in the cell nucleus.  Chromatin
fibers are packed \textit{in vivo} at a rather high density, with
volume fractions not dissimilar to those in a polymer melt. However,
unlike a melt of linear chains, the
different chromosomes in the nucleus
do not intermix, but stably occupy different distinct
``territories'': the image in which each chromosome is stained with
a different color resembles a political map of some continent
\cite{Chromosome_territories_Cremer} (see also reviews
\cite{Chromosome_territories_Cremer_Review_2006,Nature_News_Views_Territories}).
This makes an obvious analogy with the melt of nonconcatenated rings;
indeed, if the rings in the simulated melt are shown in different
colors, the image of a political map emerges\cite{PhysicsToday_Image}. This strongly
suggests topological properties of chromatin chains as a likely mechanism
behind chromosome segregation, similar to segregation of nonconcatenated rings
in the melt\cite{Vettorel09b,Arsuaga_chromosomes,Stasiak_chromosomes}.
Why the topology of the chromatin fibers is restricted is a somewhat
open question, but it might be that the DNA ends are attached to the
nuclear envelope, or simply that the cell lifetime is not long
enough for reptation to develop
\cite{Sikorav_Jannink_1994,Rosa_Everaers_PLOS_2008,vettorel_kremer2010}. Territorial
segregation of chromosomes appears to be a common feature for the cells of
higher eukaryotes, including humans, but it is less well pronounced
or absent altogether in lower eukaryotes such as yeast
\cite{Nature_News_Views_Territories,HiC_analysis_Yeast_Chromosome,Zimmer_Group_Yeast}.
This observation is at least qualitatively consistent with the fact
that topological segregation of nonconcatenated loops is fully
developed only when polymers are very long.

In recent years, more detailed information on chromatin fiber
organization has become available, due to the advent of novel
experimental techniques, such as Fluorescence In Situ Hybridization FISH \cite{FISH}
and HiC, which is a generalization of 3C (Comprehensive
Chromosome Capture)\cite{3C} and 4C (the same on a
Chip)\cite{4C}. These methods allow one to probe the large scale features of
the chromatin fiber fold in a single chromosome. In particular, the
HiC study of human genome folding
\cite{Science_2009_crumpled_globule} revealed the signature of a
fractal folding pattern which was explicitly associated with the
crumpled globule organization of DNA originally predicted on purely
theoretical grounds \cite{Grosberg_Rabin_Havlin_Neer_1993}.
Specifically, the measurement indicated that the loop factor $P(s)$
(the probability that two loci of genetic distance $s$ base-pairs
apart will be found spatially next to each other in chromosome)
scales as $P(s) \sim s^{-\gamma}$ over the interval of $s$, roughly,
$0.5 \ \mathrm{Mbp} \lesssim s \lesssim 7 \ \mathrm{Mbp}$, where the
power $\gamma$ is 1 or slightly higher.  This scaling appears to be
consistent with the crumpled globule model
\cite{Rosa_Everaers_Looping2010}.  There are also many more
indications of self-similarity, or scale-invariance, in the
chromatin folding (see recent review
\cite{Chromosome_Fractal_Review_Bancaud} and references therein).
Current biological theories about these issues also actively
involve dynamical aspects, how this presumably crumpled
self-similar structure can move to perform its functions. To this
end, there are some observations regarding the structural difference
between the active part of the chromatin fiber, currently
transcribed, and the more densely packed, non-transcribing
heterochromatin \cite{Science_2009_crumpled_globule}. Without going
into further detail, suffice it to say for the purposes of the
present paper that chromosome studies necessitate the further and
deeper understanding of the melt of nonconcatenated rings, both in
equilibrium and in dynamics, as a basic elementary model.

Recent theories for melts of nonconcatenated rings or
the globule of a single unknotted ring have been dominated by the idea
that the topological constraints imposed on a given ring by the
surrounding ones, as well as the constraints imposed by the
different parts of the same ring on each other, can be treated using
a sort of a ``topological mean field'', namely, by considering a
single ring immersed in a lattice of immobile topological obstacles
\cite{Cates_Deutsch_1986,Khokhlov_Nechaev_1985,Rubinstein_PRL_1986,Nechaev_Semenov_Koleva_1987,Obukhov_Rubinstein_Duke_PRL_1994},
with none of the obstacles piercing through the ring (such that the
ring is nonconcatenated with the whole lattice). The conformation of
a ring in such a lattice represents a lattice animal with every bond
traversed by the polymer twice (in opposite directions).  We
emphasize that the structure of branches in the lattice animal is
completely annealed. In
the off-lattice case, a similar annealed randomly-branched
structure is expected, in which the chain follows itself twice along
every branch only down to a certain scale $d$, similar to the
distance between obstacles or the tube diameter in reptation theory.
Such structure is reasonable as long as $l_k \ll d \ll \left\langle R_g \right\rangle$, where $l_k$
is the Kuhn length and $\left\langle R_g \right\rangle$ is the
average gyration radius of the loop. Based on
the lattice animal model, Cates and Deutsch
\cite{Cates_Deutsch_1986} modified Flory's argument and arrived at the
conjecture that $\left\langle R_g(N) \right\rangle \sim N^{2/5}$ for the
nonconcatenated unknotted rings in the melt.  An alternative theory,
also based on the lattice-of-obstacles picture, but consistent with
the crumpled globule model for a single ring chain \cite{Grosberg_Nechaev_Shakhnovich_1988}, 
suggested that rings
in the melt are compact objects in the sense that $\left\langle R_g(N) \right\rangle
\sim N^{1/3}$.
Lacking any systematic theoretical approaches and
based on purely heuristic arguments, this controversy has remained unresolved.

Experimental efforts to synthesize melts of pure rings have suffered
from difficulties in purification and polydispersity. Early studies
considered polystyrene and polybutadiene ring polymers synthesized
in both good and theta solvent
conditions.\cite{Hild83,Roovers83,Hild87,Roovers88} However, due to
the presence of linear chain contaminants and the possible presence
of knotted conformations, these works have been received with
skepticism. More recently ring polymers were isolated from linear
chains having roughly the same molecular weight with a very high
degree of separation.\cite{Pasch,Chang00} The concentration of
linear chains is estimated to be less than one chain in a thousand.
For these systems, however, only dynamical properties have
been reported.
Kapnistos et al. have made stress relaxation measurements for these
samples as well as samples that were contaminated systematically
with linear chains.\cite{Rubinstein_Nature_2008} The authors claim
that the viscoelastic response of the system was significantly
affected already at concentrations of linear contaminants well
below the overlap concentration. The effect of
linear contaminants will be investigated in a separate
paper\cite{Halverson10}. Nam et al. have synthesized and investigated
melts of low-molecular-weight polydimethylsiloxane rings\cite{nam_2009}. Synthesizing and isolating pure,
monodisperse, unknotted and nonconcatenated rings remains a major 
challenge and thus computer simulations of perfectly controlled
systems can be very helpful to learn more about these melts.

The early theoretical investigations were followed by a series of
computational studies.
\cite{Muller96,Brown98a,MullerBarrat00,Brown98b,Muller00,Brown01,Suzuki08,Vettorel09a,Vettorel09b,Suzuki09}
In the first work
\cite{Muller96}, the bond fluctuation model (BFM)~\cite{Carmesin88}
was employed, and rings up to $N=512$ monomers were examined at the
density of $0.5$ (in units of occupied lattice sites).  The first
results appeared to support the Cates and Deutsch
conjecture\cite{Cates_Deutsch_1986}, that $\left\langle R_g \right\rangle \sim N^{2/5}$.
However, the entanglement length for linear BFM chains at a density
of $0.5$ is known to be fairly large, $N_{e,\mathrm{linear}} >
70$.  Therefore, the rings examined in this work are still fairly short, less
than $7 N_{e,\mathrm{linear}}$, they cannot be expected to be ``asymptotic''.
In the second
study from the same group\cite{Muller00} the chain length
was extended to $N=1024$ and also stiffer chains were considered,
which at the same density have a smaller
entanglement length.
Although a crossover to $\left\langle R_g \right\rangle \sim N^{1/3}$ was observed,
the ultimate scaling could not be settled because the
overall structure of the stiff ring polymers
did not resemble a collapsed globule, being
quite extended to accommodate the higher bending energies.
More recently some of us studied a melt of nonconcatenated
flexible rings on a simple cubic lattice for chain lengths
up to $N=10000$\cite{Vettorel09b}. This was achieved by a special Monte
Carlo method which included nonlocal moves to speed-up the
relaxation\cite{Vettorel09a}. While the simulations were started from qualitatively
different initial conformations (collapsed globules, extended double
stranded paths and growing chains) the equilibrium properties were found to be the same.
For large rings, Vettorel et al. \cite{Vettorel09b} found
the $\left\langle R_g^2 \right\rangle^{1/2} \sim N^{1/3}$ scaling with
a wide crossover which can be
interpreted as a $2/5$ intermediate ``asymptotics''. 
Suzuki et al.\cite{Suzuki09} have also reproduced the
$1/3$ scaling regime. All
these works have left not only several static properties but
essentially all dynamic properties unexplored.

In the first of two papers, we investigate the static
properties of ring polymer melts using molecular dynamics (MD)
simulation. To our knowledge the only previous MD studies of ring
polymer melts are by Hur et al. \cite{Hur06} and Tsolou et al. \cite{Tsolou10} who considered
polyethylene for short chain lengths (i.e., $N\lesssim
3N_{e,\mathrm{linear}}$), and Hur et al. \cite{Yoon2010} for chain lengths up to about $11N_{e,\mathrm{linear}}$, where $N_{e,\mathrm{linear}}$ is the
entanglement molecular weight of the linear polymer system.
Here we consider ring and linear melts composed of much longer
chains. The
number of monomers per ring ranges from 100 to 1600 or $3N_{e,\mathrm{linear}}$ to $57N_{e,\mathrm{linear}}$
The linear chain melts were simulated using the same model as the rings with chain
lengths varying from 100 to 800 monomers per chain. Such highly entangled melts have very long
relaxation times and a highly parallelized simulation code was used.
For the longest rings as many as 2048 IBM Blue Gene/P cores were used for a single simulation.

The simulation model is presented in Section \ref{proto}. This is followed by a description
of the melt preparation and the simulation protocol in Section \ref{sim_meth}. The static
properties for the ring and linear melts are reported in Section \ref {statics_section}
followed by a comprehensive discussion of the results in Section \ref{discuss}, while
the key findings of
the work and future challenges are presented in the Conclusions section. Dynamic properties
based on the same simulation runs are presented in the subsequent paper\cite{halverson_part2}.

\section{Simulation Methodology}
\label{sec:2}
\subsection{Model}
\label{proto}
Based on the Kremer-Grest (KG) model \cite{Kurt90}, all beads
interact via a shifted Lennard-Jones potential (WCA or Weeks-Chandler-Anderson potential)
(Eq.~\ref{eq:pot1}) with a cutoff of $r_c=2^{1/6}~\sigma$. Covalently bonded monomers
between nearest neighbor beads along the chain also interact via the finitely extensible nonlinear
elastic (FENE) potential (Eq.~\ref{eq:pot2}).
Chain stiffness is introduced by an angular
potential (Eq.~\ref{eq:pot3}), where $\theta_{i}$ is the
angle between bonds connecting beads $i-1$, $i$ and $i$, $i+1$.\cite{Everaers04}
For all ring and linear polymer simulations reported in this work:
$k=30~\epsilon /\sigma^{2}$, $R_{0}=1.5~\sigma$ and $k_{\theta}=1.5~\epsilon$.

\begin{eqnarray}
\frac{U(r_{ij})}{4\epsilon} = \begin{cases}
         \left[\left(\frac{\sigma}{r_{ij}}\right)^{12}-\left(\frac{\sigma}{r_{ij}}\right)^{6}\right]+\frac{1}{4},
         & r_{ij} \leq 2^{1/6}~\sigma \\
         0, & r_{ij} >2^{1/6}~\sigma
\end{cases}
\label{eq:pot1}
\end{eqnarray}

\begin{eqnarray}
U(r_{ij}) = \begin{cases}
         -0.5kR_{0}^{2} \ln[1-(r_{ij}/R_{0})^{2}],     & r_{ij} < R_{0} \\
         \infty,     & r_{ij} \geq R_{0}
\end{cases}
\label{eq:pot2}
\end{eqnarray}

\begin{eqnarray}
U_{\mathrm{angle}}(\theta_{i})=k_{\theta}[1-\cos(\theta_{i}-\pi)].
\label{eq:pot3}
\end{eqnarray}

\noindent
The natural time scale of Eq.~\ref{eq:pot1} is
$\tau=\sigma\sqrt{m/\epsilon}$, where $m$ is the mass of a monomer.
For the simulations we use $T=1.0~\epsilon/k_B$ and
$\rho=0.85~\sigma^{-3}$. This model has been frequently used for
simulation studies of polymer melts and solutions, so that one can
refer to ample information throughout the discussion of the present,
new results. The linear chain data presented in this work is all
new as well. The ring melts were simulated using 200 polymers of length 100 to
1600 monomers per chain. The linear systems were composed of $2500$ chains of length $N=100$,
250 chains of $N=200$, and 400 chains of length $N=400$ and 800
(cf. Table \ref{table1}). An important parameter for our discussion is
the entanglement length for linear polymers in a melt. For the
model employed here this parameter was determined \cite{Everaers04} to be

\begin{equation}
N_{e,\mathrm{linear}}=28 \pm 1 \ . \label{eq:N_e_for_our_model}
\end{equation}

Comparing $N_{e,\mathrm{linear}}$ for different simulation models
and different chemical species allows one to compare equivalent chain lengths
from rather different studies.

\begin{table}[ht]
\begin{center}   
\begin{tabular}{c|ccc|ccc}
\hline                    
& \multicolumn{3}{c|}{Rings} & \multicolumn{3}{c}{Linear} \\
\hline                                                      
$N$ & $M$ & $L/\sigma$ & $\tau_R/10^5~\tau$ & $M$ & $L/\sigma$ & $\tau_R/10^5~\tau$ \\
\hline                                                                                
100  & 200 & 28.655 & 0.10 & 2500 & 66.503 & 0.38 \\                                  
200  & 200 & 36.103 & 0.39 &  250 & 38.891 & 1.57 \\                                  
400  & 200 & 45.487 & 1.59 &  400 & 57.287 & 6.36 \\                                  
800  & 200 & 57.310 & 6.29 &  800 & 72.176 & 25.2 \\                                  
1600 & 200 & 72.207 & 25.2 &   -- &   --   & --   \\                                  
\hline                                                                                
\end{tabular}                                                                         
\end{center}                                                                          
\caption{System parameters for the ring and linear melt simulations. All simulations were performed
with $k_BT=\epsilon$ and $\rho \sigma^3 = 0.85$ in a cubic box of side $L$. The number of monomers per chain is $N$ and the number of chains is $M$. The Rouse times \cite{Doi86,Tsolou10} are $\tau_{R,\mathrm{linear}} = N \langle R_{e,\mathrm{linear}}^2 \rangle \zeta / 3 \pi^2 k_BT$ and $\tau_{R,\mathrm{rings}}=\tau_{R,\mathrm{linear}}/4$. The friction coefficient, $\zeta$, was estimated from the initial mean-square displacement data\cite{Puetz} of the inner monomers of the $N=100$ linear system to be $43~\tau^{-1}$. It is found to be 25\% lower if the initial ring data are used.}                                                                                    
\label{table1}                                                                                                                        
\end{table}

\subsection{Preparation and Production Runs}
\label{sim_meth}

The linear chain melts were prepared following the procedure outlined in Auhl et al.\cite{Auhl}
Initially the chains are grown as random walks and placed randomly in the
simulation cell without regard to monomers overlapping. The overlaps are removed
using a soft potential whose strength is increased slowly over a few hundred thousand time
steps. The Lennard-Jones potential is then turned on. The system is further equilibrated using
the double-bridging algorithm in which new bonds
are formed across a pair of chains, creating two new chains each substantially
different from the original. This combination of methods has proven to be a very efficient approach to
equilibrate long chain melts.\cite{Auhl}

For ring melts the situation is more complex, as we do not have well known target conformations. However, from the earlier lattice studies we have some insight about the general conformations \cite{Vettorel09a}. The initial configuration for each ring melt simulation was prepared via the following steps \cite{vettorel_kremer2010}. Rings were constructed with their centers at the sites of a simple cubic lattice. The monomers of each chain were
created on a circular path in a plane with a random orientation. By
choosing the lattice constant to be larger than the ring diameter this procedure ensures that the rings are unknotted and nonconcatenated. An intrachain attractive interaction was then
used to collapse the rings. This was done by increasing the
cutoff distance in Eq.~\ref{eq:pot1} to $r_c=2.5~\sigma$. The interchain interactions
remain purely repulsive during this phase. For the $N=1600$ system, an additional force proportional to the distance between a monomer and the center-of-mass of the chain, was added in order to speed-up the initial contraction.

The density of the system was initially low because the rings were widely spaced.
A short NPT simulation for $500~\tau$ was carried out to increase the density of the system to the target value ($\rho=0.85~\sigma^{-3}$). During this phase the applied pressure was slowly increased and the intrachain attraction was kept on. Once the target density had been reached the NPT ensemble was switched to the NVT ensemble. The attractive part of the intrachain bead-bead interaction was switched off and the simulations were continued up to about three times the Rouse time of each system. The Rouse
times for the ring systems are given in Table \ref{table1}. During this equilibration phase the rings diffused a distance of at least their own diameter. Taking the typical radius of gyration, in units of the entanglement length, the conformations then compare well to our previous lattice Monte Carlo results on very long ring polymers\cite{Vettorel09b}. In the course of the following very long simulations, we also carefully monitored possible changes in the instantaneous averages of the spanning diameters and radii of gyration of the rings in order to observe any drift pointing towards poor equilibration. Within the accuracy of our data, we did not observe any drift anymore in all data used for analysis. This is especially important for ring polymers starting with somewhat artificial initial configurations\cite{Vettorel09b}.

The production runs were carried out in the NVE ensemble with a weak coupling to a Langevin thermostat \cite{schneider, Kurt90} to maintain the temperature.
A cubic simulation box with periodic boundary conditions in all dimensions was used for each simulation. The velocity Verlet method was used to numerically integrate Newton's equations of motion. The ring simulations were performed using ESPResSo \cite{limbach} and LAMMPS \cite{plimpton} with a time step and friction coefficient of 0.01~$\tau$ and $1.0~\tau^{-1}$, respectively. LAMMPS\cite{plimpton} was used for the linear systems with a time step of 0.012~$\tau$ and a friction coefficient of $0.5~\tau^{-1}$. In computing mean-square displacements the center-of-mass drift of the system was removed. Table \ref{table1} gives the system parameters for the ring and linear simulations.

\section{Conformational and Structural Properties}
\label{statics_section}

\begin{figure}[]
\includegraphics[scale=1.0]{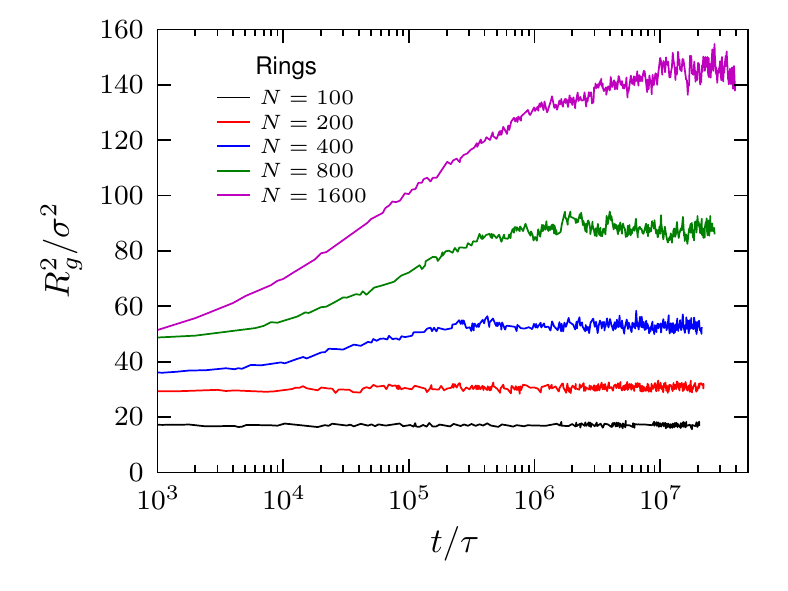}
\caption{Radius of gyration squared as a function of time for the five ring systems.}                      
\label{rg_time}
\end{figure}

The equilibration of $R_g^2$ is
shown for the ring systems in Fig. \ref{rg_time}. The larger chains
are found to increase in size before fluctuating about an average
value while $R_g^2(t)$ for the smaller chains changes very little over
the course of the simulation compared to the initial starting
conformations. Although long simulation
times are needed to equilibrate the systems, particularly
with large $N$, our runs are long enough because
 as shown in Fig. \ref{rg_time}, even for the
$N=1600$ case, the time window over which average quantities were
computed is about ten times longer than the time required for the
system to reach equilibrium. All quantities reported in
the present and subsequent paper are exclusively based on the
simulation regime where the polymers are fully equilibrated. The time window of this
regime is approximately five times the longest correlation time of the $N=1600$ ring system, as
determined by various quantities (cf. subsequent paper on dynamics\cite{halverson_part2}).
The linear systems have longer relaxation times and the simulation time of the $N=800$ linear system
did not exceed the longest relaxation time (see below).

\begin{figure}[]
\includegraphics[scale=1.0]{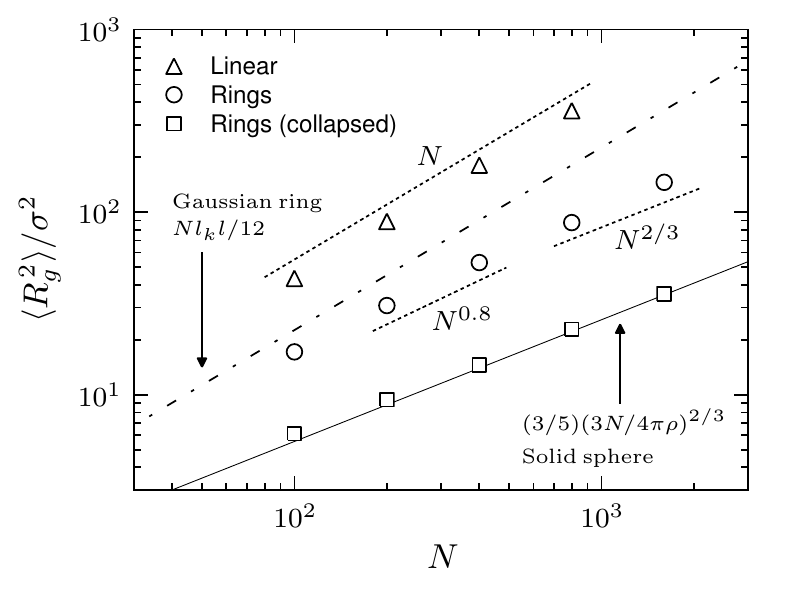}
\caption{Equilibrium values of the radius of gyration squared $\left\langle R_g^2 \right\rangle$ versus $N$ for the ring and linear systems. The dashed lines indicate the crossover and asymptotic scaling regimes of the rings. The collapsed ring data was found by simulating single rings in vacuum with $r_c=2.5~\sigma$. The solid line corresponds to a homogeneous sphere or $\left\langle R_g^2 \right\rangle = (3/5)(3N/4\pi\rho)^{2/3}$ with $\rho\sigma^3=0.85$. The error bars for all data points are smaller than the symbol sizes.}
\label{Rg_N}
\end{figure}

\begin{figure}[]
\includegraphics[scale=1.0]{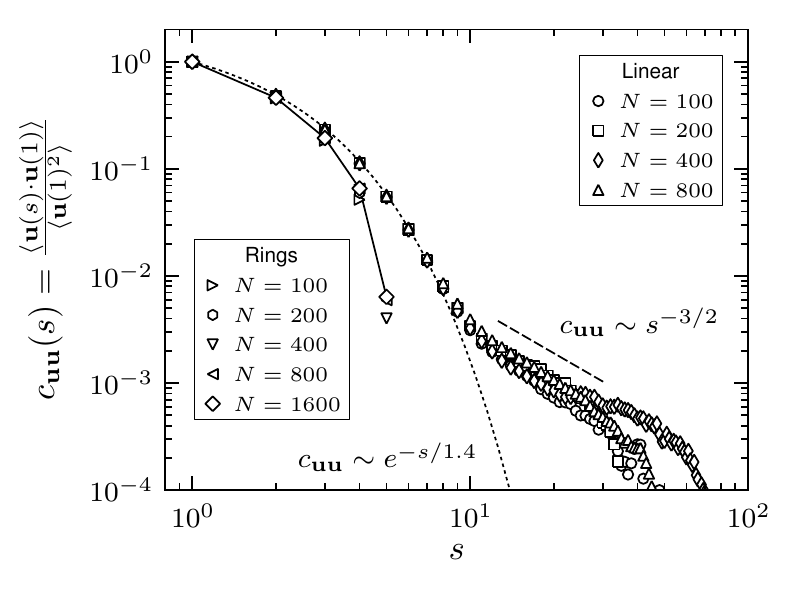}
\caption{Tangent-tangent
correlation function against contour distance between points, $s$,
in double logarithmic scale.
The correlation function is found to decay much more rapidly for the rings (solid line) than the linear chains (dotted line).}   
\label{fig:Correlations_of_Tangents}           
\end{figure}

Fig. \ref{Rg_N} shows the equilibrium average
values of  $R_g^2$
versus $N$ for the ring and linear systems. The linear chains
in the melt exhibit the expected Gaussian
behavior or $\left\langle R_g^2 \right\rangle \sim N$. We
have to emphasize that this Gaussian behavior is not altered by the
recently discovered slowly decaying tangent-tangent correlation along
the chain in the melt
\cite{Correction_to_Flory_2004,Correction_to_Flory_2007}. As Fig.
\ref{fig:Correlations_of_Tangents} demonstrates, the tangent-tangent
correlation between two points on the chain in the melt decays as
$s^{-3/2}$ with contour distance $s$, unlike exponential decay for
the ideal chain in a dilute solution. In our system, unlike the
simulations of Wittmer et al. \cite{Correction_to_Flory_2004,Correction_to_Flory_2007}, we also 
observe an initial exponential decay of the correlation at small $s$
because our chains have some significant bending rigidity (see
above). The algebraic decay of the
tangent-tangent correlation does not alter the conclusion that
linear chains in the melt are Gaussian in the sense of a normal
distribution of the end-to-end distances or simply the linear scaling of
$\left\langle R_g^2 \right\rangle$ with $N$. This is simply because the sum of $s^{-3/2}$
converges at large $s$, indicating that an effective segment comprises
only a finite number of monomers.

The ring systems show various regimes depending on their
lengths. For small $N$ they appear Gaussian. An extended crossover
regime then follows where $\left\langle R_g^2 \right\rangle \sim N^{4/5}$. Finally, the longest
chains approach a $N^{2/3}$ scaling, showing that very long ring
polymers, with $N \gtrsim 400 \approx 15
N_{e,\mathrm{linear}}$ (see Eq.~(\ref{eq:N_e_for_our_model}))
are required to approach the asymptotic regime. Exactly the same is observed for the spanning
distance squared $R_e^2(N)$ of the rings, which is defined as the
mean-square distance between monomers $1$
and $N/2+1$ (or $k$ and $k+N/2$ with any $k$) in the ring (cf. Table
\ref{table2}). These results are in agreement with previous
simulations \cite{Vettorel09b}.

\begin{figure}[]
\includegraphics[scale=1.0]{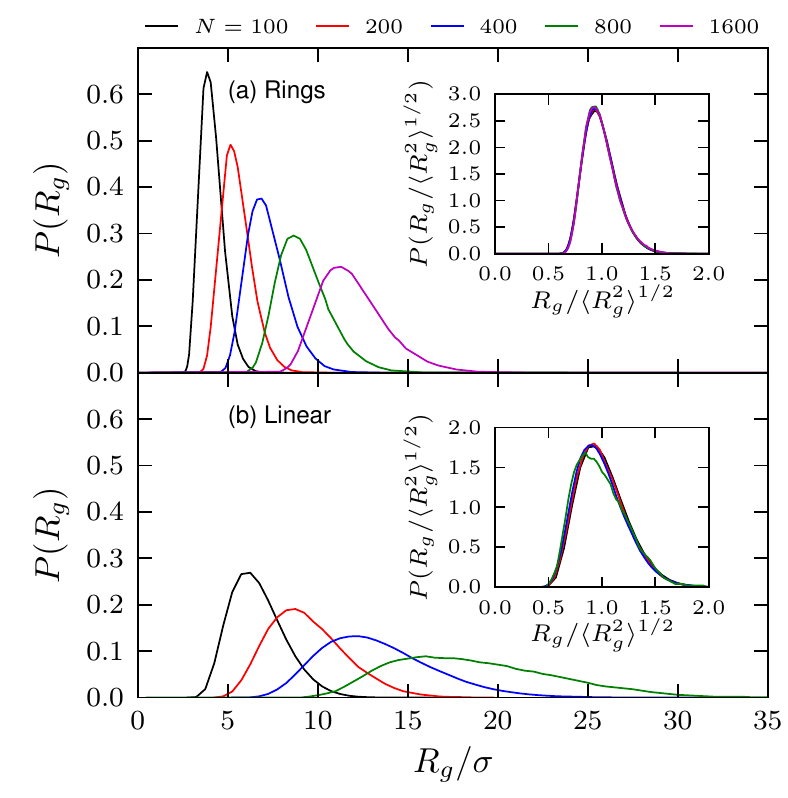}
\caption{Probability distribution function of the radius of gyration for the (a) rings and (b) linear chains. The insets show the probability distribution function of $R_g$ normalized by its root-mean-square value.}
\label{Rg_dist}
\end{figure}

The equilibrium fluctuations of the gyration radius are given
in Fig. \ref{Rg_dist} for each of the ring
and linear systems. In this paragraph we reserve
the notation $R_g$ for the instantaneous (fluctuating) value of the
gyration radius, while the mean-square gyration radius is written as
$\left\langle R_g^2 \right\rangle$.
The distributions become wider and asymmetrical with increasing
$N$, however they follow a perfect scaling when normalized
by $\left\langle R_g^2 \right\rangle^{1/2}$.
The insets to Fig. \ref{Rg_dist} give the
probability distribution functions of
$R_g$ normalized by the respective
root-mean-square values for the (a) rings and (b) linear chains. All five
data sets for rings of different lengths are found to
collapse nearly perfectly onto a single curve.  For the
linear chains, while data for $N=100$, $200$, and $400$
collapse about as well as for the rings, the distribution
for the $N=800$ case slightly deviates from the others. This is indicative of
a system that may not have been completely equilibrated despite an extremely
long simulation time.
Some properties
are more sensitive to improper equilibration than others. For this
work only the shear relaxation modulus, which is discussed in
the subsequent paper, is affected by the lack of
equilibration for the $N=800$ linear case.
Our data suggest that all other linear and ring systems are
properly equilibrated.

\begin{table*}[ht]
\begin{center}
\begin{tabular}{|c|cccc|cccc|}
\hline
& \multicolumn{4}{c|}{Rings} & \multicolumn{4}{c}{Linear} \\
\hline                                                      
$N$ & $\left\langle R_g^{2} \right\rangle / \sigma^2$ & $\left\langle R_e^{2} \right\rangle / \sigma^2$ &
 $\langle\lambda_1\rangle/\langle\lambda_3\rangle$ & $\langle\lambda_2\rangle/\langle\lambda_3\rangle$ & 
 $\left\langle R_g^{2} \right\rangle / \sigma^2$ & $\left\langle R_e^{2} \right\rangle / \sigma^2$ &     
 $\langle\lambda_1\rangle/\langle\lambda_3\rangle$ & $\langle\lambda_2\rangle/\langle\lambda_3\rangle$ \\
\hline                                                                                                   
 100 &  17.2 (0.4) &  50.8 (1.5) & 6.4 & 2.3 &  43.4 (1.2) &  263.8 (1.6) & 12.9 & 2.8 \\                
 200 &  30.8 (0.7) &  88.8 (2.7) & 5.9 & 2.2 &  88.9 (1.2) &  538.9 (1.6) & 12.6 & 2.8 \\                
 400 &  52.9 (1.2) & 149.4 (4.8) & 5.5 & 2.1 & 180.8 (1.3) & 1095.3 (1.6) & 12.3 & 2.8 \\                
 800 &  87.6 (1.9) & 242.4 (7.3) & 5.3 & 2.0 & 359.1 (1.3) & 2164.9 (1.7) & 11.9 & 2.7 \\                
1600 & 145.6 (2.8) & 401.7 (10.7)& 5.2 & 2.0 &      --       &      --        &  --   & --   \\          
\hline                                                                                                   
\end{tabular}                                                                                            
\end{center}                                                                                             
\caption{                                                                                                
Size and shape of the ring and linear chains. $\left\langle R_g^2 \right\rangle$ is the mean-square gyration radius
and $\left\langle R_e^2 \right\rangle$ is the mean-square spanning distance between monomers $N/2$ apart for the rings
and the mean-square end-to-end distance for the linear chains. Error bars for these quantities are indicated in
parentheses. The average eigenvalues of the gyration tensor are                                                                       
arranged as $\langle\lambda_1\rangle \geq \langle\lambda_2\rangle \geq \langle\lambda_3\rangle$                                       
with $\left\langle R_g^2 \right\rangle=\sum \langle \lambda_i \rangle$.                                                               
}                                                                                                                                     
\label{table2}                                                                                                                        
\end{table*}

The average values of $R_g^2$ and $R_e^2$ as well as the ratios of the
average eigenvalues of the gyration tensor for the rings and linear chains
are shown in Table \ref{table2}. Not only are the rings
significantly more compact than the linear polymers, they also
display a quite different shape. For the linear systems
the mean-square end-to-end distance is found to be related to
the mean-square gyration radius via $\left\langle R_e^2 \right\rangle /
\left\langle R_g^2 \right\rangle = 6.00 \pm 0.02$, 
in excellent agreement with the expected value of $6$ for
ideal chains; this confirms once again that linear chains
in the melt are Gaussian. For the rings, a similar role is played by the ratio of
the mean-square spanning distance $\left\langle R_e^2 \right\rangle$ (mean-square value of the vector
connecting two monomers $N/2$ apart along the ring) to the mean-square gyration
radius $\left\langle R_g^2 \right\rangle$; this ratio for the rings varies
from $\left\langle R_e^2 \right\rangle / \left\langle R_g^2 \right\rangle
\approx 3.0$ for $N=100$ to about $2.75$ for both the $N=800$ and $1600$, indicating
that with the approach of $\left\langle R_g^2 \right\rangle \sim N^{2/3}$ this ratio is close to asymptotic. This is
supported by plotting $\left\langle R_e^2 \right\rangle / \left\langle R_g^2 \right\rangle$ versus
$1/N$ and extrapolating to infinite chain lengths where one finds a value of $2.75\pm0.03$.
For Gaussian rings
$\left\langle R_e^2 \right\rangle / \left\langle R_g^2 \right\rangle = 3$;
while short rings are close to Gaussian by this measure, the long ones deviate
noticeably from Gaussian statistics. Both ring and linear
architectures are found to have shapes that resemble prolate
ellipsoids with the linear chains being
significantly more elongated. As $N$ increases the shape of the
rings becomes a little closer to spherical
with the eigenvalue ratios of the largest chains being 5.2:2.0:1 (to be
contrasted to the eigenvalue ratios of roughly 12:3:1 of a Gaussian
coil).\cite{Kurt90,Kremer1992_shape}
This behavior agrees well with the lattice simulations of
Vettorel et al. \cite{Vettorel09b} when the difference in
$N_{e,\mathrm{linear}}$ between the two models is taken into
account.

Further information about rings and chain shapes can be
obtained from higher moments. Specifically, we can define the following series of characteristic lengths:

\begin{equation} R_g^{(2m)} = \left[ \frac{1}{N}\sum_{i=1}^{N} \left\langle \left(
\boldsymbol{r}_{i} - \boldsymbol{r}_{\mathrm{CM}}\right)^{2m} \right\rangle
\right]^{1/2m} \ . \label{eq:generalized_moment_definition}
\end{equation}

In this formula, $\boldsymbol{r}_{i}$ is the position vector of
monomer $i$, while $\boldsymbol{r}_{\mathrm{CM}}$ is the position vector
of the mass center of the entire coil.  Obviously, $\left\langle R_g^2 \right\rangle^{1/2} \equiv
R_g^{(2)}$ ($m=1$). Moments with higher $m$ are sensitive to
extending ``peninsulas'' of the structure.  Table
\ref{tab:Higher_Moments} shows moments up to $m=5$ for both the rings and linear
systems.  For the linear chains, the results are in nearly
perfect agreement with the theoretical values for Gaussian coils,
confirming yet again that chains in the melt are practically
Gaussian.  By contrast, the situation with the rings is much less
simple: despite the (approximate) $N^{1/3}$ scaling of $\left\langle R_g \right\rangle$, 
characteristic of a
compact object, the values of the higher moments are nowhere near those
for the hard sphere, and, surprisingly, not very far from
theoretical results for Gaussian rings.  This indicates that
although rings in the melt are overall relatively compact, they do
have many loops which protrude quite far from the center-of-mass of the chain.

\begin{figure*}[]
\includegraphics[scale=1.0]{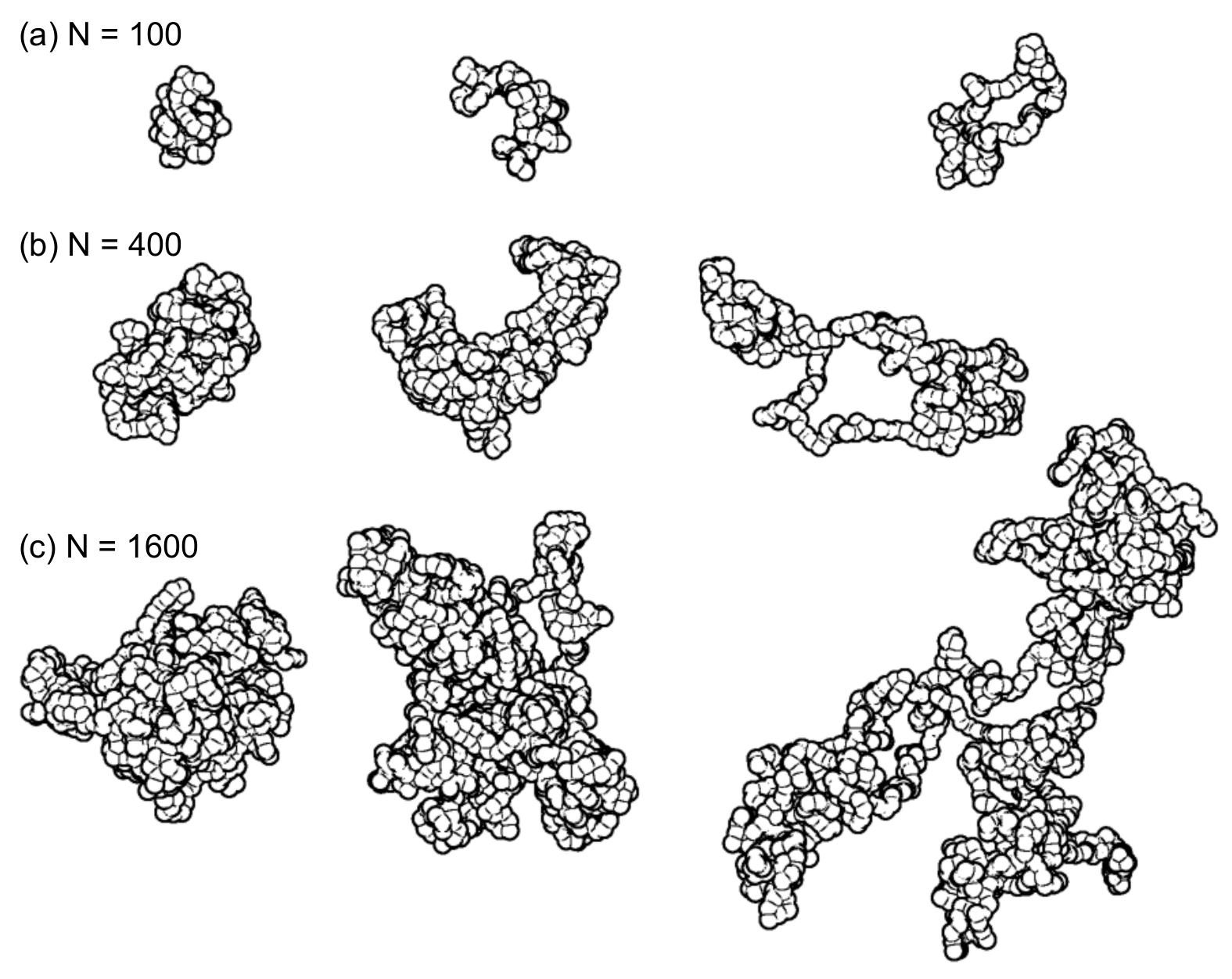}
\caption{Individual chain snapshots for three of the ring systems. The left and right columns show small and large chains, respectively, while the middle column shows a chain having nearly the average value of $R_g^2$.}                      
\label{nvt_snapshots}                              
\end{figure*}

\begin{table*}[ptb]
\label{tab:regimes}
\begin{tabular}{|c|c|c|c|c|c|c|c|c|c|c|c|c|c|}
\hline                                        
& & \multicolumn{5}{c|}{Linear} & \multicolumn{6}{c|}{Rings} & Solid \\
\cline{1-13}                                                           
& $m$ & $N=100$ & 200 & 400 & 800 & Gaussian & $N=100$ & 200 & 400 & 800 & 1600 & Gaussian & Sphere\\
\hline $R_g^{(4)}/R_g^{(2)}$ & $2$ & $1.18$ & $1.18$ & $1.19$ & $1.18$ & $1.19$ & $1.13$ & $1.14$ & $1.14$ & $1.15$ & $1.15$ & $1.14$ & $1.04$ \\                                                                                                                  
\hline $R_g^{(6)}/R_g^{(2)}$ & $3$ & $1.34$ & $1.35$ & $1.36$ & $1.36$ & $1.36$ & $1.24$ & $1.26$ & $1.27$ & $1.28$ & $1.28$ & $1.25$ & $1.07$ \\                                                                                                                  
\hline $R_g^{(8)}/R_g^{(2)}$ & $4$ & $1.48$ & $1.50$ & $1.51$ & $1.52$ & $1.53$ & $1.35$ & $1.36$ & $1.38$ & $1.40$ & $1.41$ & $1.36$ & $1.10$ \\                                                                                                                  
\hline $R_g^{(10)}/R_g^{(2)}$ & $5$ & $1.61$ & $1.64$ & $1.65$ & $1.67$ & $1.67$ & $1.44$ & $1.46$ & $1.49$ & $1.51$ & $1.52$ & $1.46$ & $1.11$ \\                                                                                                                 
\hline                                                                                                                                
\end{tabular}                                                                                                                         
\caption{\label{tab:Higher_Moments}Average higher moments of the single-chain distance distribution for the linear and ring systems as computed by Eq.~\ref{eq:generalized_moment_definition}. The values for a Gaussian linear chain and Gaussian ring are also given for comparison with the MD results. For a solid sphere $R_g^{(2m)}/R_g^{(2)}=(5/3)^{1/2}[3/(2m+3)]^{1/2m}$. Note that $\langle R_g^2 \rangle^{1/2} = R_g^{(2)}$.}                                                                                                                        
\end{table*}

\begin{figure*}[]
\includegraphics[scale=1.0]{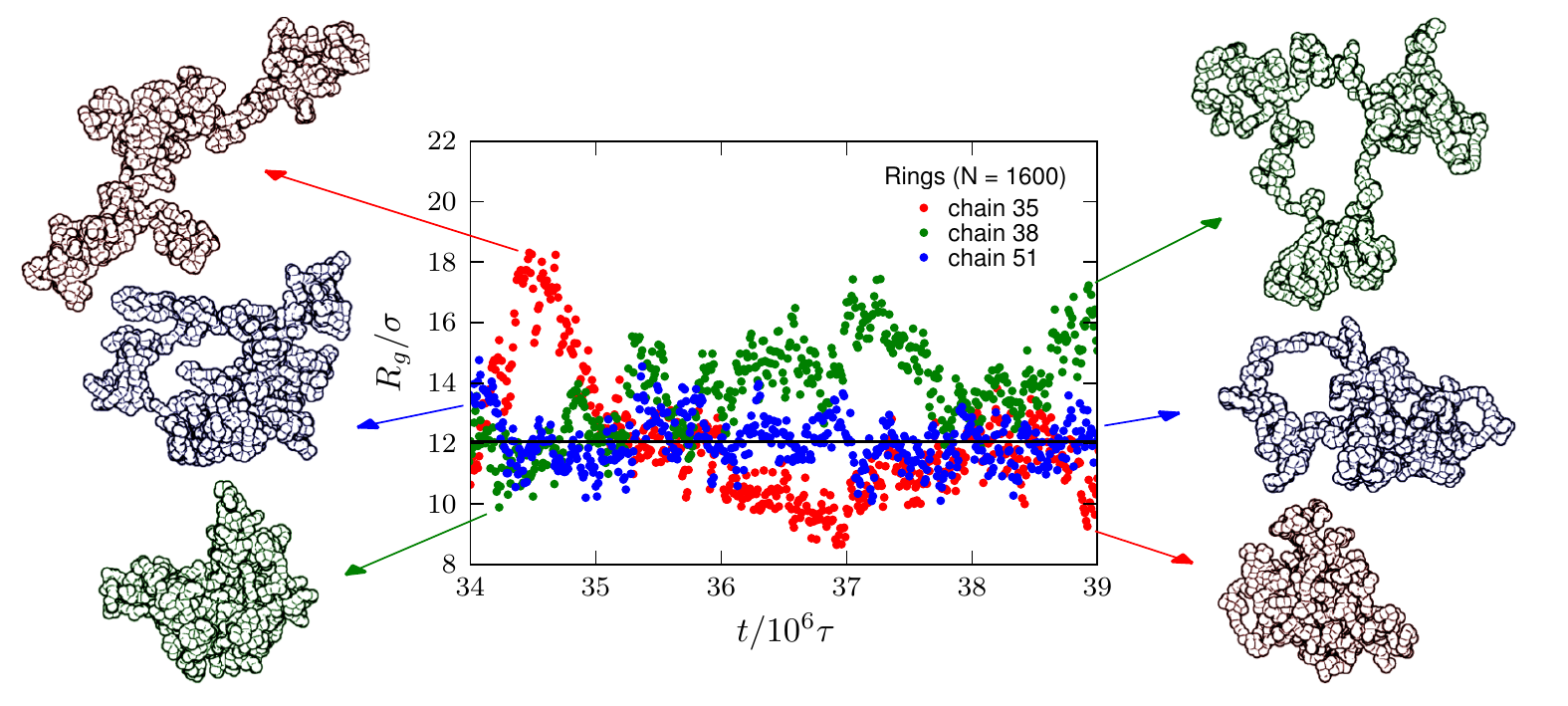}
\caption{Radius of gyration for three individual chains from the $N=1600$ ring melt simulation as a function of time along with instantaneous configurations. The horizontal line indicates the value of $\left\langle R_g^2 \right\rangle^{1/2}$.}
\label{rg_t_graph}
\end{figure*}

To illustrate this Fig. \ref{nvt_snapshots} shows three individual ring snapshots for
each of the $N=100$, 400 and 1600 systems. Chains are taken from the
wings and center of the distribution of $R_g$. The extensions
increase from left to right with the chain in the middle chosen to
have approximately $\left\langle R_g^2 \right\rangle$. The smallest chains for
each system are found to be collapsed while the largest are expanded
and exhibit voids which are occupied by neighboring chains. This
last point is confirmed by a primitive path
analysis in the subsequent paper\cite{halverson_part2} and self-density data which is presented below.
The conformations in
the left and center columns of Fig. \ref{nvt_snapshots} are representative of the crumpled globule where
each chain has a globule-like core surrounded by protrusions of size ${\cal O}(N_e)$.
The variety of conformations in Fig. \ref{nvt_snapshots} raises the point of 
whether there is a significant entropy
barrier between the quasi-collapsed state of a ring and an extended
conformation, which requires significant ring-ring interpenetration
without concatenation. To illustrate that this barrier is very small
or even nonexistent (as suggested by the unimodal distributions in
Fig. \ref{Rg_dist}a), Fig. \ref{rg_t_graph} shows $R_g(t)$ and
snapshots for three individual polymers from the $N=1600$ ring
simulation. Chain 35 has a highly extended conformation with a large
value of $R_g$ around $t=34.5 \times 10^6~\tau$, however, $4.5
\times 10^6~\tau $ later the chain is found to assume a less
extended conformation with a corresponding reduction in $R_g$ of
50\%. A similar behavior is found for chain 38 while the size of
chain 51 remains at approximately the root-mean-square value (as indicated by
the horizontal line) over the entire time window.

\begin{figure}[]
\includegraphics[scale=1.0]{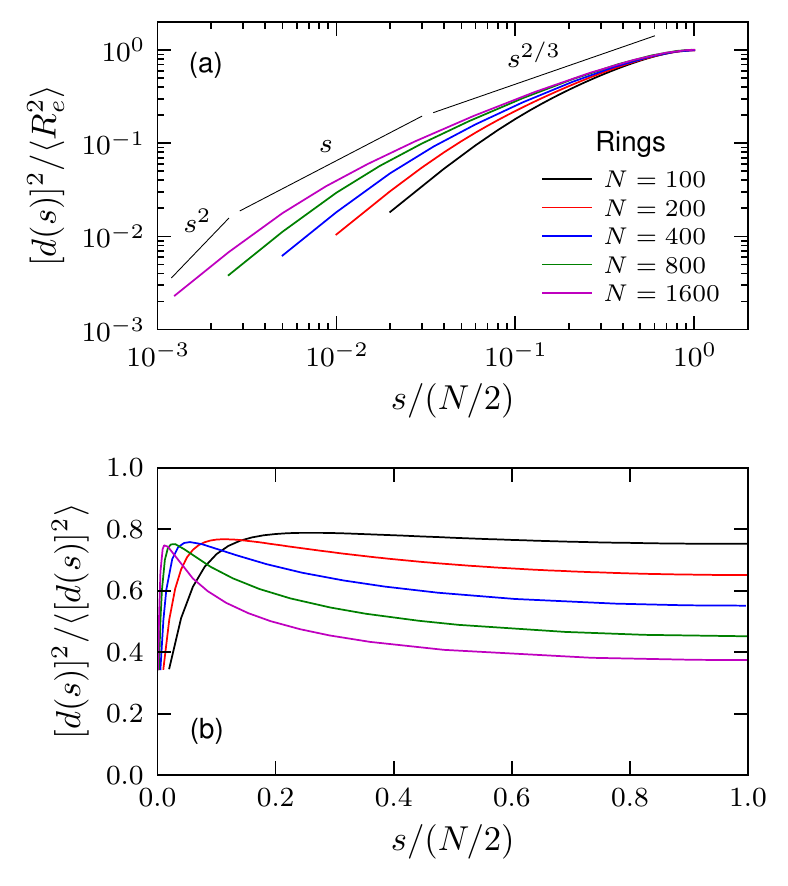}
\caption{Internal structure of the rings as characterized by the mean-square internal distances normalized by (a) the mean-square spanning distance and (b) the scaling for a Gaussian ring or $\left\langle [d(s)]^2 \right\rangle = l_k l/[1/s+1/(N-s)]$, where $s$ is the number of bonds separating two monomers and $1 \leq s \leq N/2$. Normalizing by $\left\langle R_e^2 \right\rangle$ in (a) causes all five data sets to coincide at $s=N/2$.}
\label{id_n}
\end{figure}

The internal structure of the chains may be characterized by the
mean-square internal distances. Let $[d(s)]^2$ denote the average squared distance
between two monomers separated by $s$ bonds. This
quantity normalized by $\langle R_e^2 \rangle$ is shown in Fig.
\ref{id_n}a. The local rigidity of the chains leads to an
approximate $s^2$ scaling for short separations. An intermediate
regime with a linear scaling in $s$ is found for $l_k/d(s=1) \ll s
\ll N/2$, where $l_k$ is the Kuhn length which for our model\cite{Everaers04} is
$2.79~\sigma$. The $[d(s)]^2 \sim s$ behavior is
found for linear chains where it is also the asymptotic scaling. For
the rings at very large separations a $s^{2/3}$ regime is found
before $[d(s)]^2$ becomes independent of $s$ as $s$ approaches half of the chain. In Fig.
\ref{id_n}b the mean-square internal distances are normalized by those of a
Gaussian ring\cite{zimm_stockmayer} or $\left\langle [d(s)]^2 \right\rangle=l_k l
/[1/s+1/(N-s)]$, where $l$ is the average bond length. Although the overall shape of the
$d(s)$ dependence is quite reminiscent of that of a Gaussian ring,
it is definitely not the same. Of course, this is not surprising
because unlike linear chains, rings in the melt are not Gaussian.

\begin{figure}[]
\includegraphics[scale=1.0]{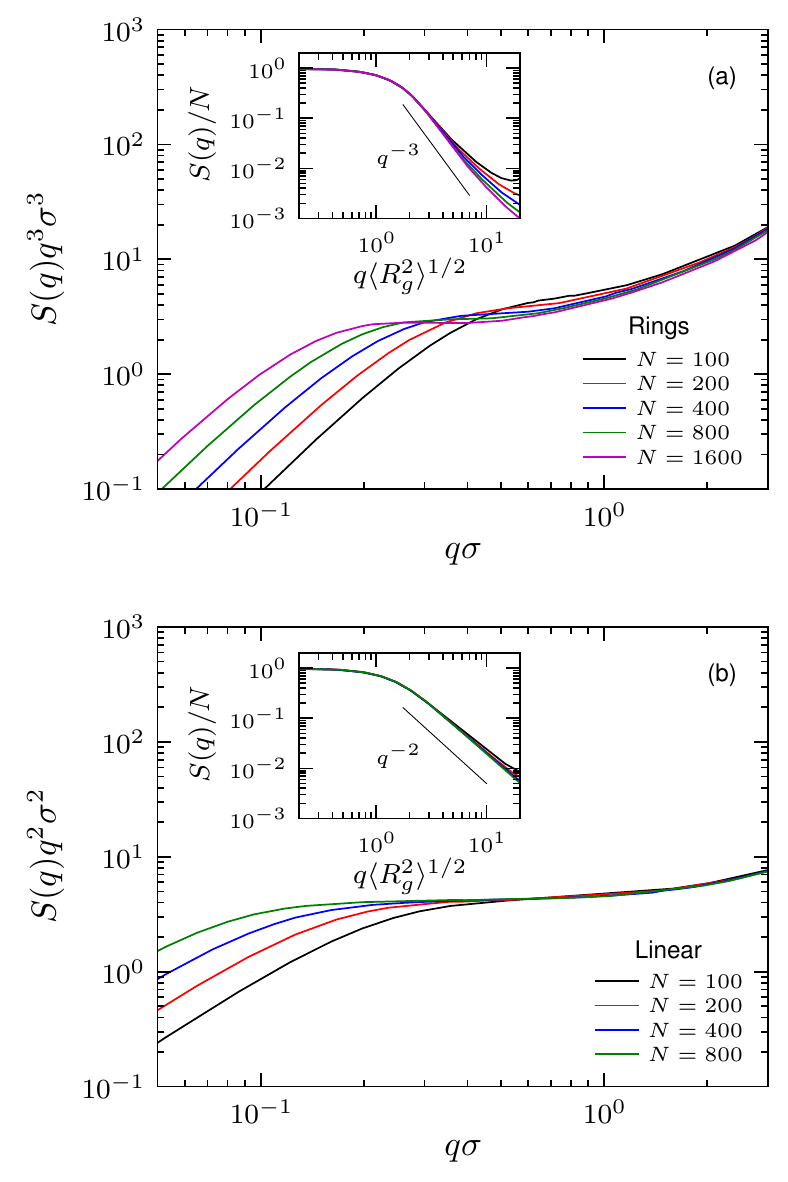}
\caption{The single-chain static structure factor multiplied by (a) $q^3$ for the rings and (b) $q^2$ (Kratky plot) for the linear chains.}
\label{struxqsquare_q}
\end{figure}

The static structure function of the individual rings multiplied by
$q^3$ is shown in Fig. \ref{struxqsquare_q}.
For $2\pi/\left\langle R_g^2 \right\rangle^{1/2} < q \ll 2\pi/l_k$ one
expects $S(q) \sim q^{-1/\nu}$ for self-similar structures. While this
scaling is almost perfectly fulfilled for linear chains with $\nu=1/2$, the
situation is more complex for the rings. If the large rings behaved
like homogeneous, compact objects then one would expect $S(q)$ to be
dominated by Porod scattering\cite{Grest87,Grest89} or $S(q) \sim
q^{-4}$. This is certainly not the case here.
The best fit to $S(q) \sim q^{-x}$ in this intermediate self-similar scaling regime
is $x=3.2 \pm 0.1$. This will be explained in a more general context in the discussion section.
The inset of Fig. \ref{struxqsquare_q}a shows $S(q)/N$ versus $q \left\langle R_g^2 \right\rangle^{1/2}$.
For large values of $q$ one finds a
gradual increase in $S(q)q^3$ indicating a conformation between an
overall collapsed, but very irregularly and self-similarly shaped
object. All systems show a qualitatively self-similar behavior over
a wide range of $q$ indicating that the general chain shape is the
same between cases. The shift in the overall amplitude of
$S(q)q^3$ in the intermediate regime, however, indicates that, unlike for
linear chains, there is some deviation from the ideal fractal object
structure, i.e. a weak ring length dependence of the density
of scatterers.

\begin{figure}[]
\includegraphics[scale=1.0]{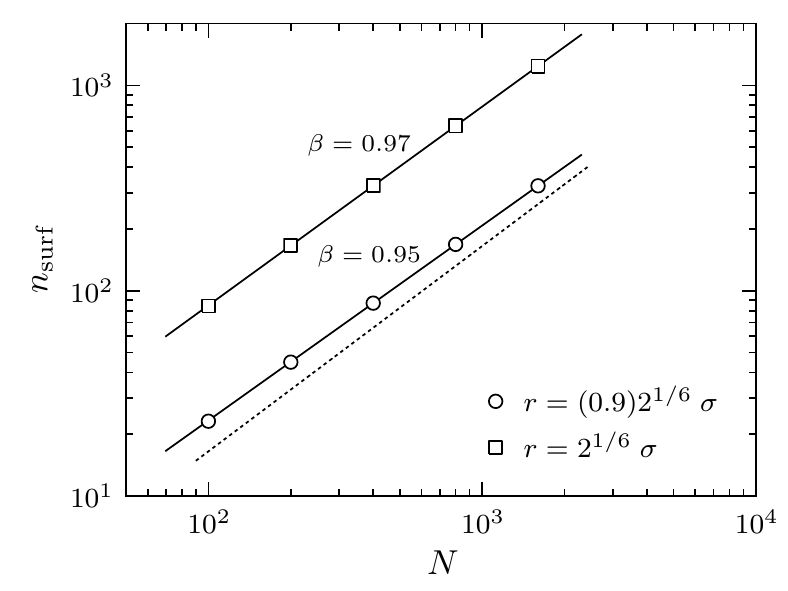}
\caption{Number of surface monomers per chain versus $N$ for two choices of the cutoff distance as indicated. The error in values of
$n_{\mathrm{surf}}$ is approximately 0.5\%, significantly smaller than the symbol size, suggesting that the observed exponent can be distinguished from 1 (dashed line).}                      
\label{surface_atoms}                              
\end{figure}

Taking the information revealed by Figs. \ref{id_n} and
\ref{struxqsquare_q}, the apparently changing exponents from
$\nu=1/2$ for very short rings to $\nu=1/3$ for the long ones can be
seen as a crossover into the asymptotic regime without significant
qualitative influence on the internal structure. Actually, one would
expect each ring to have a fairly large number of surface beads
($n_{\mathrm{surf}}$), where a
bead is a surface bead if it is within a distance $r$ of at least one bead
of another chain. For random
walks, i.e. linear melts, ``all'' beads (${\cal
O}(N)$) are surface beads, while for a compact
sphere only ${\cal O}(N^{2/3})$ are surface beads. For the rings we find
$n_{\mathrm{surf}} \sim N^{\beta}$, $\beta \approx 0.95$ (see Fig. \ref{surface_atoms}). This
power law covers
the whole range from $N=100$ to 1600 without any sign of a
crossover. The observed exponent of $0.95$ is very close to
unity, so one is inclined to suspect that in fact $n_{\mathrm{surf}}
\sim N^1$, however, the data do suggest that this power is
slightly below unity and there are also theoretical grounds to
expect it to be smaller than 1.0 (see Appendix).

\begin{figure}[]
\includegraphics[scale=1.0]{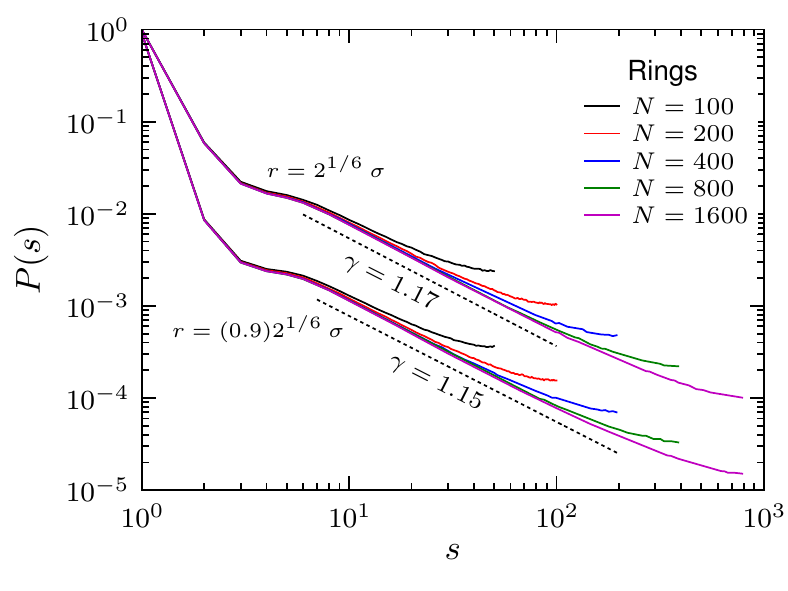}
\caption{$P(s)$ is the probability of two monomers of the same chain being separated by $s$ bonds and a distance of less than $r$.}
\label{gamma_exponent}
\end{figure}

The internal structure of the chains can also be characterized by $P(s)$, which is the
probability of two monomers of the same chain being separated
by $s$ bonds and a distance of less than $r$. In Fig. \ref{gamma_exponent}, $P(s)$
is shown for the ring melts for the same two choices of the cutoff distance as in Fig. \ref{surface_atoms}. For the larger
rings one finds a power law of $P(s)\sim s^{-\gamma}$ for intermediate values of $s$. 
The value of $\gamma$ is found to decrease weakly with increasing $s$. For the larger
cutoff distance and $N=1600$, $\gamma$ varies from 1.17 when the data
is fit to values of $s$ between $30-50$ to 1.00 for values between $110-130$. The same
trend is seen for the smaller value of $r$.

To finally confirm that the rings do not assume double-stranded conformations we
computed the probability distribution function of the angle
formed by vectors $\boldsymbol{a}=\boldsymbol{r}_{i+n}-\boldsymbol{r}_i$ and
$\boldsymbol{b}=\boldsymbol{r}_{j+n}-\boldsymbol{r}_j$, where $\boldsymbol{r}_k$ is the
position vector of monomer $k$ and $j>i$.
The calculation
was performed only when $|\boldsymbol{r}_i-\boldsymbol{r}_j|$ was less than a
distance, $r$, which was taken as a multiple of the excluded-volume
interaction cutoff and covered a range from nearest-neighbor separations to approximately
the tube diameter ($2\left\langle R_g^2(N_e) \right\rangle^{1/2} \approx 7~\sigma$) as based on
the linear chain system. For all combinations of $n=1,2,3$ and $r/r_c=1,2,3,4,5$ the distribution
function was found to be consistent with the vectors having random
orientations confirming that rings in the melt are not double-stranded. Such
a finding casts severe doubt on approaches which are based on lattice animal arguments.

\begin{figure}[]
\includegraphics[scale=1.0]{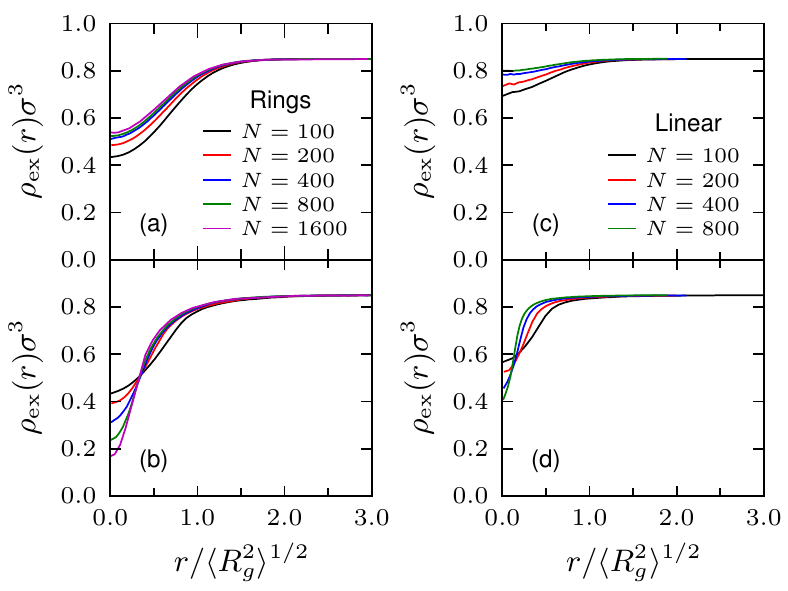}
\caption{Self-excluded monomer density profiles for the rings and linear systems. For (a) and (c), $r=0$ corresponds to the center-of-mass of the central chain while for (b) and (d) it corresponds to the center of maximum density.}
\label{se_rdf}
\end{figure}

The entropy loss associated with the nonconcatenation constraint
suggests that a ring melt
should have a higher pressure than a linear
melt at the same density and temperature. This is supported by the
finding that a single ring in a linear melt (where there is no
concatenation constraint) has a larger size than in a melt of
rings.\cite{Iyer07} We can estimate the pressure contribution due to the restricted topology of nonconcatenated rings
by following the ideas of Ref.~36. According to that work, entropy loss due to
topological constraints is of the order of the number of rings contacting a given ring, the
quantity called $K_1$ below. That corresponds to a contribution to the pressure which is of the order of
$k_BT K_1/Nv$, where $Nv$ is the volume of a ring. Given that $k_BT=\epsilon$ (see Simulation Methodology
section above), and $v \approx \sigma^3$, we arrive at the estimate of the pressure as $K_1/N$ in the units of
$\epsilon/\sigma^3$. Given that $K_1$ is less than 20 (see below), we arrive for $N=800$ at the
pressure contribution of less than $0.025~\epsilon/\sigma^3$.
While this estimate is open for criticism, since it is based on the method of Ref.~36 which
does not produce the correct scaling exponent in
$\left\langle R_g \right\rangle \sim N^{\nu}$,
the deviation is not so large that the total pressure would be significantly changed.

The above estimate of the topological contribution to the pressure is compared to the overall pressure which was measured by MD simulation and found to be $P = 4.99 \pm 0.02~\epsilon/\sigma^3$ for $N=800$. Thus, the estimated topological contribution is two orders of magnitude smaller than the pressure and is roughly the same magnitude as the error bars of our measurements. It is therefore not surprising that we obtained an indistinguishable pressure of $5.00 \pm 0.01~\epsilon/\sigma^3$ for the melt of linear chains at the same conditions. Further work is required to detect the topological contribution to the pressure and examine its exact scaling behavior.

\begin{figure}[]
\includegraphics[scale=1.0]{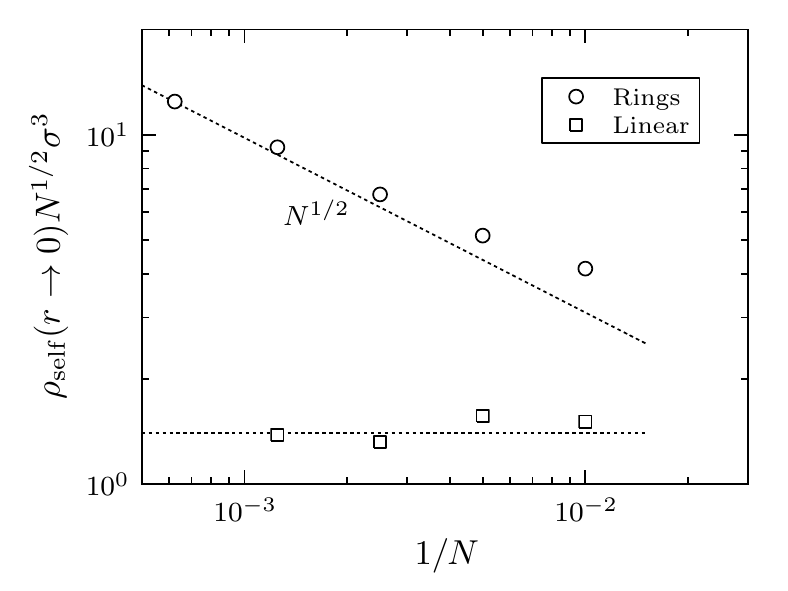}
\caption{Log-log plot of the self-density as $r \rightarrow 0$ multiplied by $N^{1/2}$ versus inverse chain length
for the rings and linear polymers. $\rho_{\mathrm{self}}$ around the center-of-mass for linear polymers
vanishes with $N^{-1/2}$ while for the rings it approaches a chain length independent constant of
$\rho_{\mathrm{self}} (N \rightarrow \infty) \approx 0.31~\sigma^{-3}$. Note
that $\rho_{\mathrm{self}}(r)=\rho-\rho_{\mathrm{ex}}(r)$.}
\label{rho_center}
\end{figure}

The overall properties of a single ring in the melt display the
general scaling of a compact object. However, the
conformations are open enough in order to allow for a significant
interpenetration of rings. Such rather extended shapes require that
the depth of the correlation hole, which measures the self-density
of the chains, be significantly deeper than for linear melts. For
linear polymers the volume of each chain is shared by $N^{1/2}$
other chains, resulting in a vanishing self-density of the chains,
i.e. $\rho_{\mathrm{self}} \sim N^{-1/2}$. For a compact object,
independent of the overall bead density, $\rho_{\mathrm{self}}$
eventually has to become independent of $N$. In Fig. \ref{se_rdf}
the self-excluded density profile is shown for the rings and linear chains. This
profile is computed in the standard way for Figs. \ref{se_rdf}a,c except
that the origin is
taken as the center-of-mass of the central chain and the monomers of
the central chain are ignored in the calculation.
For each system
the magnitude of $\rho_{\mathrm{ex}}$ in the vicinity of $r=0$ does
not vanish. This arises from the mutual interpenetration of chains
and is more pronounced for the linear.
$\rho_{\mathrm{ex}}(r \rightarrow 0)$ is found to
approach a constant for the longer rings ($N \geq 400$).
For Figs. \ref{se_rdf}b,d
the center of maximum density of the central chain is used. This is found by
a binning procedure with a bin size of $5~\sigma$.
The self-density as $r \rightarrow 0$ is shown in Fig. \ref{rho_center} where it can be seen
that this quantity for the rings approaches a constant as $N \rightarrow \infty$.

The behavior of $\rho_{\mathrm{ex}}(r)$ is consistent with the
observed decrease in self-density with increasing $N$, if estimated
from $\left\langle R_g^2 \right\rangle^{1/2}$. With the
self-density being $\rho_{\mathrm{self}}=3N / 4 \pi
\left\langle R_g^2 \right\rangle^{3/2}$ (which assumes a homogeneous sphere), we find
$\rho_{\mathrm{self}}=0.34~\sigma^{-3}$ for $N=100$ while for the
large rings ($N \geq 400$) this value drops to around 0.23. Since
the bulk density is $0.85~\sigma^{-3}$ this implies that roughly
$70\%$ of the monomers within 
$\left\langle R_g^2 \right\rangle^{1/2}$ of the center-of-mass of the long rings come from
other chains. This, in combination with $\nu=1/3$ means that the
number of different rings in direct contact with a given ring should
asymptotically become independent of $N$.

\begin{figure}[]
\includegraphics[scale=1.0]{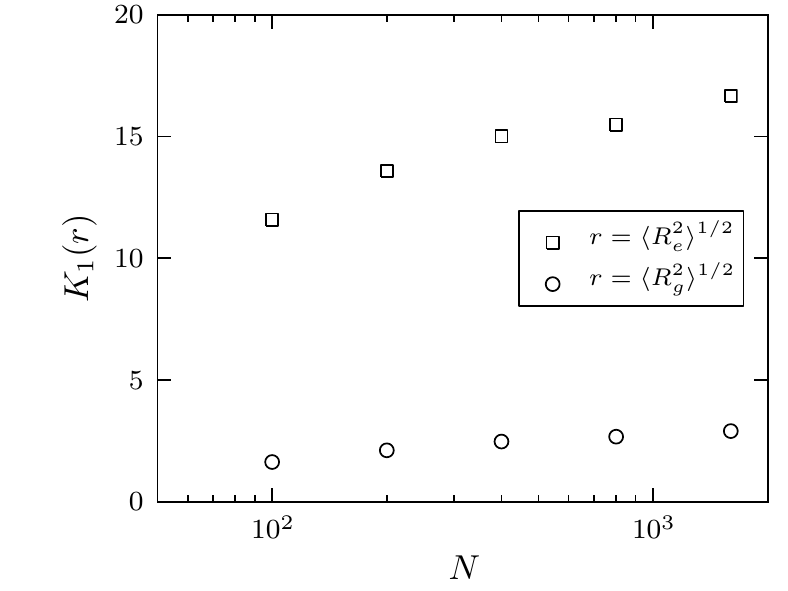}
\caption{Number of neighboring chains versus $N$. $K_1(r)$ is the average number of chains within a
center-of-mass separation distance $r$.}
\label{nN}
\end{figure}

The structure of a polymer melt is partly determined by the number
of neighbors per chain. 
Let $K_1(r)$ denote the average number of chains
whose center-of-mass is within a distance
$r$ of the center-of-mass of a
reference chain. 
This quantity is shown in Fig. \ref{nN} as a
function of $N$ for the rings. As chain length
increases $K_1(r)$ approaches a constant for both
$r=\left\langle R_g^2 \right\rangle^{1/2}$ and $r=\left\langle R_e^2 \right\rangle^{1/2}$, which is in agreement with our arguments above (i.e., as $N \rightarrow \infty$, $\nu \rightarrow 1/3$).
This is to be contrasted with the fact that $K_1(\left\langle R_g^2 \right\rangle^{1/2})$ for the melt of linear chains grows as $N^{1/2}$. This agrees with the finding that the rings in the melt,
although collapsed overall, do have significant protrusions into one
another. The lattice simulations of Vettorel et al.
\cite{Vettorel09b} gave a limiting value of $K_1(\left\langle R_g \right\rangle)
\approx 14$ as $N \to \infty$ which agrees well with the value of
around 16 for the $N=1600$ system in the present work.
This interplay of (partial)
segregation and (partial) interpenetration is expected to play an
important role for the dynamics of the melts, which is our focus in
the following paper.

\section{Discussion}
\label{discuss}
While the linear chains show a $\left\langle R_g^2 \right\rangle \sim
N$ behavior for all chain lengths, the rings behave differently.
While the conjecture of Cates and Deutsch \cite{Cates_Deutsch_1986} based on a
free energy argument that the Flory exponent for the radius of
gyration of a ring polymer melt should be approximately $2/5$ has
reportedly been confirmed by a number of simulation
\cite{Muller96,Brown98b,Muller00,Iyer07,Vettorel09b,Suzuki09}
and theoretical \cite{Suzuki09} studies, it now appears that this
corresponds to an intermediate regime with the asymptotic Flory exponent being $1/3$
\cite{Muller00,Vettorel09b,Suzuki09}.
The present work is in
agreement with the scaling laws for these two regimes. Vettorel et
al. \cite{Vettorel09b} estimate the onset of the asymptotic regime using an
argument based on $N_{e,\mathrm{linear}}$. For the scheme where
chains form a core which squeezes out the other rings, these authors
estimate a critical ring length $N_c$ to be
\begin{eqnarray}
N_c \approx \rho (4 \pi /3) \left\langle R_{e,\mathrm{linear}}^2(N_{e,\mathrm{linear}})\right\rangle^{3/2},
\end{eqnarray}
which in our case leads to $N_c \approx 2300$. Thus it is clear that
even with our longest rings, which based on a comparison of the
entanglement length for polystyrene would correspond to a molecular
mass of more than $10^6$~g/mol, we only would observe the crossover towards
the $\left\langle R_g^2 \right\rangle \sim N^{2/3}$ scaling. While this estimate assumes that the rings
squeeze each other out more or less completely, the data show that
this is not the case. The incomplete segregation of
rings, in which the self-density $\rho_{\mathrm{self}}$ is about $1/3$
of the bulk density inside any particular coil reduces the expected
crossover chain length to $N_c \approx 700-800$ in good
agreement with our data. This is actually a lower limit for $N_c$, since the
crumpled globules are not ideal spherical objects, and it illustrates that one needs
extremely long rings to unambiguously investigate the asymptotic regime.

Although rings in the melt appear to be compact objects in
the sense that $\left\langle R_g^2 \right\rangle \sim N^{2/3}$ at sufficiently large $N$, their
compactness is of a non-trivial character.  In many cases, they
are torus (or even pretzel)-shaped; their higher moments are much larger
than those of a solid sphere; and their average self-density, although it does
not asymptotically depend on $N$, remains at roughly $\rho/3$ instead of $\rho$. These facts
indicate that different rings in the melt, although reasonably
segregated in the scaling sense, have nevertheless significant
protrusions into one another. This can also be seen in the critical
exponents describing the system.

In the familiar linear polymer coil, everything is governed by the
single Flory exponent $\nu$, which describes the coil gyration
radius. Although the rings in the melt appear to exhibit the asymptotic
scaling $\left\langle R_g^2 \right\rangle^{1/2} \sim N^{1/3}$, their
properties are not completely determined by the value of the exponent
($\nu=1/3$), at least not in a trivial way. There are two
important examples of exponents that are not directly related to $\nu$.
The first describes the number of ``surface monomers'' (see above,
Section \ref{statics_section}), i.e., monomers of a given polymer that have
contacts with other polymers: $n_{\mathrm{surf}} \sim N^{\beta}$; we
find that our simulations yield $\beta \approx 0.95$. We argue
that the same power $\beta$ should describe the number of contacts
between two crumples of $g$ monomers each, whether belonging to the
same ring or to two different rings; this number of contacts scales
as $g^{\beta}$. The second is the $\gamma$ exponent which describes the
looping probability -- the probability that two monomers of the
same ring, a distance $s$ along the chain, will be found next to
each other in space. The looping probability scales as
$s^{-\gamma}$. Understanding this exponent for rings in the melt is of particular interest because this system, as mentioned in the introduction section, is expected to capture essential features of chromatin packing in chromosomes. The exponent $\gamma$ can be measured directly for chromatin using HiC methods \cite{3C,4C,Science_2009_crumpled_globule,HiC_analysis_Yeast_Chromosome}.  We now argue that there is a general scaling relation between exponents $\beta$ and $\gamma$, namely, $\beta + \gamma = 2$ (see Appendix for derivation).  Our simulation data, presented above (see Figs. \ref{surface_atoms} and \ref{gamma_exponent}) are closer to $\beta + \gamma \approx 2.1$.  It is not completely clear to us whether this result
agrees with the theoretical prediction to within the error bars of the simulation data,
or the deviation is real. Uncertainty in the simulation data is thought to arise from finite size effects.  Interestingly, the simulation result for $\gamma \approx 1.1$ is also close to the mean field prediction $\gamma = 3 \nu$ (see Appendix) as well as to the experimental result for chromatin \cite{Science_2009_crumpled_globule} $\gamma \approx 1$.  Thus, our current understanding is that $\beta$ is slightly smaller than unity, while $\gamma$ is slightly larger than unity.  The former observation indicates that there are many contacts between crumples, but still not all of the monomers are in contact, i.e., there is some segregation between crumples; in terms of modeling the chromatin, this is consistent with the concept of chromosome territories \cite{FISH}.  The later observation is also consistent with the idea that the system is a fractal.  Indeed, a true fractal on an unlimited spectrum of scales cannot be realized with $\gamma \leq 1$ (because then the total number of contacts per monomer, proportional to $\sum_{1}^{N} s^{-\gamma}$ would diverge).  Thus, the observed values of $\beta$ and $\gamma$ are consistent with the non-trivial segregated fractal structure.

\begin{figure}[]
\includegraphics[scale=1.0]{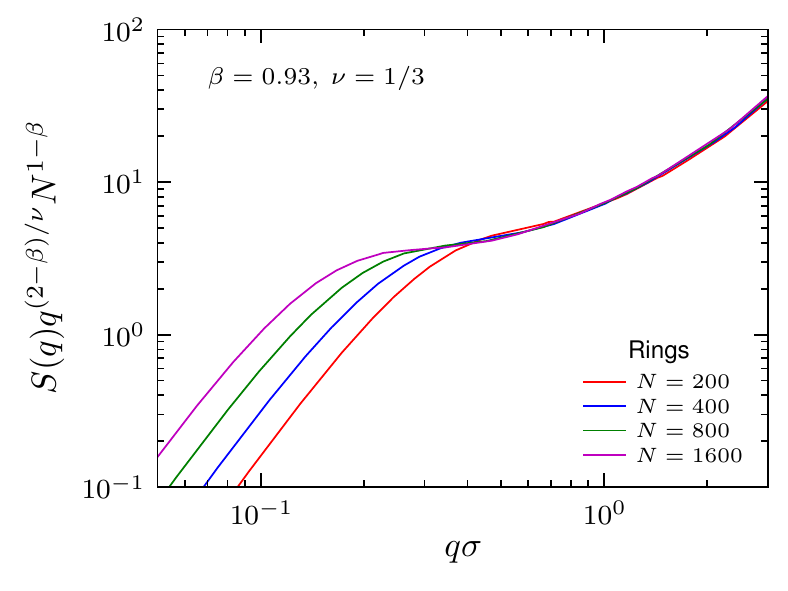}
\caption{The single-chain static structure factor multiplied by the general scaling factor $q^{(2-\beta)/\nu} N^{1-\beta}$ with $\beta=0.93$ and $\nu=1/3$ for the rings with $N \geq 200$.}
\label{sq_shura_scaling}
\end{figure}

The fact that we observe deviations from a generally expected $S(q) \sim q^{-1/\nu}$
scaling in the regime $2\pi/\left\langle R_g^2 \right\rangle^{1/2} \ll q \ll 2\pi/\sigma$ as well
as the finding $\beta < 1$ can be rationalized by
generalizing a recent investigation of 2-d polymer melts by Meyer et al.\cite{meyer10}
Their analysis is based on a rather general theoretical framework of B. Duplantier\cite{duplantier89}.
While the Duplantier theory applies in 2-d only, and no 3-d generalization is
available (precisely because of the difficulties with topological
constraints), the scaling considerations can be developed for our 3-d
system in the following way.
In
a 2-d polymer melt the chains form compact objects and partially segregate.
The situation here is
somewhat similar and we can generalize this concept to our problem of a melt
of nonconcatenated rings. In general, with $n$ scatterers (being the whole ring
or just a part):

\begin{eqnarray}
S(q) = \frac{1}{n}\left \langle \sum_{i,j=1}^n \exp [i\boldsymbol{q} \cdot (\boldsymbol{r}_i - \boldsymbol{r}_j)]\right \rangle,
\end{eqnarray}

\noindent
where one finds for $q \rightarrow 0,~S(q) \rightarrow n$. For the intermediate scaling regime,
for a fractal structure one expects $S(q) \sim (q\left\langle R_g \right\rangle)^{-x}$. These two
arguments imply $S(q) \sim n(q\left\langle R_g \right\rangle)^{-x}$ with $x$ to be determined. The
total scattering intensity $I(q)=nS(q)$ for intermediate values of $q$ can only result from the beads at the surface of the ring
polymer, as these are the only places where a scattering probe experiences a contrast.
Consequently we expect $I(q) \sim n^{\beta}/q^x$. Combining all this leads to the scaling
relation $2-x\nu=\beta$ or $x=(2-\beta)/\nu$ and a power law of

\begin{eqnarray}
S(q) \sim n^{\beta - 1}/q^{(2-\beta)/\nu}
\end{eqnarray}

\noindent
in the intermediate scattering regime. For a random walk, where ``all'' beads are
surface beads, $\beta=1$ and $\nu=1/2$, we recover the well known $q^{-2}$ power law
for $S(q)$, which also is very well reproduced by our data. For the rings the situation
is more complex. The best estimate for $\beta$ is 0.95 as shown in Fig. \ref{surface_atoms}.
This would lead to a small amplitude shift in $S(q)$ between $N=100$ and $N=1600$ of
about $10-20\%$, which indeed is observed. In addition, with $\nu=1/3$ we
find $(2-\beta)/\nu \approx 3.1-3.2$ also in good agreement with the data. Fig. \ref{sq_shura_scaling}
shows the scaling plot for $S(q)$ displaying the overall consistency of the data
as well as the result that the best fit of $\beta$ for the scattering data is 0.93, which once
again agrees with $\beta$ less than 1. Keeping in mind
that very long rings are needed to reach the asymptotic regime makes a precise
estimate of $\beta$ quite difficult. However, the overall consistency of $S(q)$ and
the independent determination of $\beta$ strongly support the general scheme and $\beta < 1$.

For linear chains longer than $N_{e,\mathrm{linear}}$, topological
constraints manifest themselves in that a ``tube defect'' (one
strand leaking out of its tube) becomes entropically unfavorable
when the involved excursion reaches a length larger than $L_e = b
\sqrt{N_{e,\mathrm{linear}}}$. In some form, similar behavior is
expected in the case of rings as well. After the initial
Gaussian-like scaling of very short chains, the lattice animal
structure then emerges as the next natural conformation. However,
this structure is restricted due to the same arguments, which make
tube leakage very unfavorable. Any extended branch of a lattice
animal would suffer from a severe entropy penalty. This restricts
the average branch extension to a volume with a radius of
approximately the tube diameter or to a chain length of about
$2N_{e,\mathrm{linear}}$.  Importantly, effective obstacles forcing
the topological constraints are imposed not only by the surrounding
chains, but also by the remote parts of the same chain.  Unlike for a melt
of linear chains, this is important for rings, because
rings in the melt are compact and, therefore, the fraction of other
pieces of the same chain among the spatial neighbors of any monomer
is much higher in the ring system than in the linear melt.  This
eventually creates the crumpled globule conformation with rather
small branches of size ${\cal O}(N_{e,\mathrm{linear}})$,
bulging in and out at any place along the contour. A related lattice
of obstacles model suggests that the nonconcatenated ring in the melt to
some extent can be thought of as an annealed branched polymer. It is
annealed in the sense that the material can move from one branch
to another in the course of thermal motion, as Milner and Newhall
have recently emphasized \cite{milner10}. The
important difference however is that, as the data nicely show, the
densely-folded regions can open up and allow for significant
interpenetration.

\section{Conclusions}
\label{sec:Conclusions}
Thus far we have presented a rather detailed study of the
static properties of a melt of nonconcatenated rings.
Even though asymptotically the ring extension follows a
$N^{1/3}$ power law, suggesting that the rings have an overall globular conformation,
their structure is far from trivial. Rings still display
significant mutual interpenetration and the entropic barrier
to form fairly large open loops is, in fact, so small that they
are frequently observed. Though overall compact this leads
to the fact that almost ${\cal O}(N)$ of the beads are surface beads. These structural peculiarities
have important consequences for dynamics, as will be shown
in the subsequent paper\cite{halverson_part2}. Though already quite extensive, the
current analysis leaves out many properties of relevance to biology, i.e.
chromatin packing. Quantities specifically motivated by biological
problems such as inter- and intra-ring contact maps will be studied
in a separate publication.

\begin{acknowledgments}
The authors are grateful to T. Vilgis, T. Vettorel and V. Harmandaris
for their comments on an early version of the manuscript. The ESPResSo
development team is acknowledged for optimizing the simulation
software on the IBM Blue Gene/P at the Rechenzentrum Garching in
M\"{u}nich, Germany. We thank Donghui Zhang for discussions and
references relating to experimental studies on cyclic polymers. This
project was in part funded by the Alexander von Humboldt Foundation
through a research grant awarded to AYG. AYG also acknowledges
the hospitality of the Aspen Center for
Physics where part of this work was done. WBL acknowledges
financial support from the Alexander von Humboldt Foundation 
and the Basic Science Research Program through the National Research Foundation
of Korea (NRF) funded by the Ministry of Education, Science and
Technology (2010-0007886). Additional
funding was provided by the Multiscale Materials Modeling (MMM) initiative of the Max Planck
Society. We thank the New Mexico Computing Application Center NMCAC
for a generous allocation of computer time. This work is supported by
the Laboratory Directed Research and Development program at Sandia
National Laboratories. Sandia is a multiprogram laboratory operated
by Sandia Corporation, a Lockheed Martin Company, for the United
States Department of Energy under Contract No. DE-AC04-94AL85000.
\end{acknowledgments}

\appendix

\section{Relation between critical exponents of surface monomers and loop factor}
\label{sec:beta_plus_gamma}

The purpose of this appendix is to derive the relation between
powers $\beta$ and $\gamma$, as mentioned in the section
\ref{discuss}.  To begin with, recall the definitions of
these powers: \begin{itemize} \item Exponent $\gamma$ describes the
loop factor, namely, the probability that two monomers a contour
distance $s$ apart meet each other in space decays with $s$ as
$s^{-\gamma}$.
\item Exponent $\beta$ determines the number of ``surface
monomers'', $n_{\mathrm{surf}} \sim N^{\beta}$.
\end{itemize}

To derive the requisite relation between $\beta$ and $\gamma$, we
start by noting that it is equally well applicable to the
self-similar internal structure of any particular ring squeezed
between other rings. In this latter case, exponent $\beta$
describes the number of monomer-monomer contacts between any two
blobs, or crumples. Specifically, if two blobs are of some $g$
monomers each, and they are in contact (in the sense that the
distance between them is about their own size), then the number of
monomer-monomer contacts between them scales as $g^{\beta}$.

With this in mind, consider a ring in terms of $N/g$ blobs, $g$
monomers each. Concentrate on pairs of blobs which are a distance $s$
monomers, or $s/g$ blobs apart, along the chain.  There are $N/g$
such pairs in the chain.  Among these pairs some fraction are in
contact. This fraction scales as a power of contour distance, i.e.
it scales as $(s/g)^{-\gamma}$; therefore, the total number of blobs in
contact is $(N/g) (s/g)^{-\gamma} = (N/s^{\gamma}) g^{\gamma - 1}$.

Now, let us consider monomer contacts instead of blob contacts.
First, monomers are not in contact if they belong to non-contacting
blobs.  Second, given that there are $g^{\beta}$ monomer contacts per pair of
contacting blobs, we can find the total number of contacts between
monomers: $(N/s^{\gamma}) g^{\gamma - 1 + \beta}$.  Third, we have
to realize that what we have counted is the number of contacting
monomer pairs which are a distance $s \pm g$ along the chain.  The
number of monomer contacts a distance exactly $s$ apart is at least $g$ times
smaller, i.e. $(N/s^{\gamma}) g^{\gamma - 2 + \beta}$.  Finally, the
number of monomer contacts cannot depend on how we counted them,
that is, it cannot depend on $g$.  Therefore, we arrive at

\begin{equation}
\gamma + \beta = 2.
\label{eq:relation_between_indices}
\end{equation}

As an example, consider a Gaussian polymer coil in 3-d.  In this case,
$\gamma = 3/2$ and $\beta = 1/2$, so formula
(\ref{eq:relation_between_indices}) works.  In general, for the
Gaussian coil in $d$ dimensions, we have $\beta = 2-d/2$ and $\gamma
= d/2$, so again it works.

As mentioned in the main text, $\gamma > 1$ is required for
a mathematically rigorous fractal structure, which is self-similar
over an infinite range of scales.  This implies then that $\beta <
1$; in this sense, different crumples of one chain or different
rings must be segregated from each other, meaning that the number of
``surface monomers'' scales at least slower than the total $N$.

Although we have established the general relation between exponents
$\beta$ and $\gamma$, theoretical determination of any one of them
remains an open challenge.  Mean field arguments suggest that for
any fractal conformation with $r(s) \sim s^{\nu}$ in 3-d, the
probability for two monomers to meet within a small volume $v$
should go as $v/r^{3}(s) \sim s^{-3 \nu}$, which means $\gamma = 3
\nu$. This gives the familiar result $s^{-3/2}$ for the
Gaussian coil, while for a crumpled globule this yields $s^{-1}$,
i.e., an impossible $\gamma =1$. Of course, this only means the
inapplicability of the mean field argument: monomers $s$ apart
remain correlated in some yet unknown way instead of being
independently distributed over the volume of the order of
$r^{3}(s)$.  Nevertheless, it is worth noting that $\gamma$ was
found experimentally to be very close to unity for human chromosomes
\cite{Science_2009_crumpled_globule}, while $\beta$ is found very
close to unity in the present work. Clearly, $\gamma=1$ is
impossible for a rigorous mathematical fractal, but it might be very
close to unity for a real physical system which is only
approximately self-similar even if over a very wide range of scales.
It remains to be understood how all these facts are connected to
each other.


%

\end{document}